\documentclass[12pt,preprint]{aastex}
\usepackage{amsmath}

\shorttitle{MW \& RRL}
\shortauthors{Ablimit at el.}

\begin{document}


\title{The Density Profile and Kinematics of the Milky Way with RR Lyrae Stars}
\author{Iminhaji Ablimit\altaffilmark{1,2} and
Gang Zhao\altaffilmark{1} }
\altaffiltext{1}{Key Laboratory of Optical Astronomy, National Astronomical Observatories,
Chinese Academy of Sciences, Beijing 100012, China; iminhaji@nao.cas.cn}
\altaffiltext{2}{Department of Astronomy, Kyoto University, Kitashirakawa-Oiwake-cho, Sakyo-ku, Kyoto 606-8502, Japan; iminhaji@kusastro.kyoto-u.ac.jp}


\begin{abstract}
Most of known RR Lyraes are type ab RR Lyraes (RRLab),
and they are the excellent tool to map the Milky Way and its substructures.
We find that 1148 RRLab stars determined by Drake et al.(2013) have been observed by spectroscopic surveys of SDSS and LAMOST. We derived radial velocity dispersion, circular velocity and mass profile from 860 halo tracers in our paper I. Here, we present the stellar densities and radial velocity distributions of thick disk and halo of the Milky Way. The 288 RRLab stars located in the thick disk have the mean metallicity of [Fe/H]$=-1.02$. Three thick disk tracers have the radial velocity lower than 215 km $\rm s^{-1}$.
With 860 halo tracers which have a mean metallicity of [Fe/H]$=-1.33$, we find a double power-law of $n(r) \propto r^{-2.8}$ and $n(r) \propto r^{-4.8}$ with a break distance of 21 kpc to express the halo stellar density profile. The radial velocity dispersion at 50 kpc is around 78 km $\rm s^{-1}$.

\end{abstract}


\section{Introduction}


The recent photometric and spectroscopic surveys have been providing information of stars to measure
the stellar number density profile and kinematics
of the Galactic disk and halo (e.g., Bland-Hawthorn \& Gerhard 2016), and the stellar density profile is
the one of the key issues for knowing the nature of the Milky Way. The Galactic thick disk is
different from the Galactic thin disk by its unique chemistry, older age and higher elevation (e.g.
Bensby 2014; Hawkins et al. 2015). It has been showed that an edge in the stellar disk could be in the
range $R_{\rm GC} = 10-15$ kpc from the surveys data (Habing 1988; Minniti et al. 2011).

The Milky Way's stellar halo is an important component to understand our Galaxy's formation history.
Many observational studies have claimed the broken power-law slope of space density distributions with
RR Lyraes (RRLs), globular clusters and blue horizontal branch stars so on (e.g., Saha 1985; Wetterer \& McGraw 1996; Miceli et al. 2008; Bell et al. 2008; Watkins et al. 2009; Sear et al. 2010, 2011; Deason et al. 2011). The theoretical study of Font et al. (2011) predict the broken power-law slope of the mass-density profile. The study of Deason et al. (2013) always found the broken stellar halo profile of the Galaxy, and they theoretically explained the origin of the break radius in the RRL profiles in terms of the apocentric distance of the satellites that were accreted. More recently, Iorio et al. (2017) demonstrated the properties of the Galactic stellar halo with RRL in the Gaia data, and they showed that the inner halo ($R_{\rm GC} < 28$ kpc) stellar density profile is well approximated by a single power-law with exponent $\alpha = -2.96$. Cohen et al. (2017) presented RRLs determined in the Palomar Transient Facility database, and derived the stellar density profile with the power-law index of -4 for the outer halo of the Milky Way.


RRL is a very good standard candle because of the narrow luminosity-metallicity-relation in the visual band and period-luminosity-metallicity relations in the near-infrared wavelengths. Besides, most of RRL stars are type ab RRLs (RRLab), and they are old ($>10$ Gyr) with low metallicity, so that they are mainly distributed in the bulge, thick disk and halo (Smith 1995; Demarque et al. 2000).

In this paper, we investigate density profiles of the Galactic thick disk and halo by using 1148 RRLab variable stars presented in our previous paper (Ablimit \& Zhao 2017, hereafter paper I).
We briefly describe our RRL sample and methods in \S 2.
The kinematic and density information of the thick disk and halo are
discussed in \S 3. In the last section (\S 4), the conclusions are given.

\section{The sample and methods}
\label{sec:model}

The RRLab sample of this work were selected from 12227 RRLab stars of Catalina
Surveys Data Release 1 (Drake et al. 2013a), and the cross matches were made with
Sloan Digital Sky Survey (SDSS) spectroscopic data release 8 (DR8) and
Large sky Area Multi-Object fiber Spectroscopic Telescope
(LAMOST) DR4 within an angular distance of 3 arcsecond.
LAMOST is a Chinese national scientific research facility operated
by National Astronomical Observatories, Chinese Academy of Sciences
(Zhao et al. 2006, 2012; Cui et al. 2012). We found 797 LAMOST matched RRLab
stars and 351 SDSS matched RRLab stars with the corrected radial velocities
(uncertainty $<15$ $\rm km\,\rm s^{-1}$) and reliable metallicities (in the region of
Galactic centric distance $\leq 50$ kpc, for more details see paper I).

An absolute magnitude-metallicity relation has been
adopted for distance determination (Sandage 1981). The method of Chaboyer (1999) and Cacciari
\& Clementini (2003) is
used in this work as,
\begin{equation}
M_{\rm V} = (0.23 \pm 0.04)([\rm Fe/H] + 1.5) + (0.59 \pm 0.03),
\end{equation}
where [Fe/H] is the metallicity of an RR Lyrae
star. We find our overall uncertainties is around 0.15 mag from the uncertainties from the photometric calibration and the variations in metallicity and uncertainty in RRab absolute magnitudes (also see Dambis et al. (2013) for the uncertainty in absolute magnitude). The uncertainties of $\sim 7\%$ in distances are expected by our overall uncertainties of $M_{\rm V}$. The heliocentric $d$ and Galactocentric distances $R_{\rm GC}$ can be derived from the equations,
\begin{equation}
d = 10^{(<V> - M_{\rm V} + 5)/5}\, {\rm kpc},
\end{equation}
where $<V>$ average magnitudes were corrected for interstellar medium extinction
using Schlegel et al. (1998) reddening maps, and
\begin{equation}
R_{\rm GC} = (R_\odot - d{\rm cos}\,b\, {\rm cos}\,l)^2 + d^2{\rm cos}^2\,b\, {\rm sin}^2\,l +
d^2{\rm sin}^2\,b \, {\rm kpc},
\end{equation}
where $R_\odot$, $l$ and $b$ are the distance from the sun to the Galactic center (8.33 kpc in this work, see Gillessen et al. 2009), Galactic longitude and latitude of the stars, respectively.

We obtain the fundamental (Galactic) standard of rest (FSR) of stars by using the heliocentric radial velocities ($V_{\rm h}$, the corrected ones:see our paper one) and the solar peculiar motion  of (U, V, W) =
(11.1, 12, 7.2) km ${\rm s}^{-1}$ (Binney \& Dehnen 2010) which are defined in a right-handed Galactic system with U
pointing toward the Galactic center, V in the direction of rotation,
and W toward the north Galactic pole. The value of $235\pm7$ km ${\rm s}^{-1}$ is taken for the local standard of rest ($\rm{V}_{\rm lsr}$, Reid at al. 2014) in the equation below,
\begin{equation}
V_{\rm FRS} = V_{\rm h} + {\rm U} {\rm cos}\, b\, {\rm cos}\, l + ({\rm V} + {\rm V}_{\rm lsr})
{\rm cos}\, b\, {\rm sin}\, l + {\rm W} {\rm sin}\, b.
\end{equation}

For deriving the spatial density of our RRLab sample, we followed the density calculation
as a function of Galactocentric distance described by Wetterer \& McGraw (1996), as
\begin{equation}
{\rho}({R_{\rm GC}}) = \frac{1}{4\pi {R^2_{\rm GC}} f({R_{\rm GC}})}\frac{dN}{dR}
\end{equation}
$N$ is the number of RRLab as a function of distance and $f({R_{\rm GC}})$ is the fraction
of the total halo volume at $R_{\rm GC}$ sampled by the survey. The efficiency or completeness
of sampling (selection process) is a way to achieve each $f({R_{\rm GC}})$ for each individual field.
Drake et al. (2013a) discussed the Catalina Surveys efficiency of RRL sampling, and their
Figure 13 showed the detection completeness as a function of magnitude. We followed
Drake et al. (2013a) for $f({R_{\rm GC}})$ by adopting completeness as 70\% for $V < 17.5$
mag and it is gradually reach to 0\% from $V = 17.5$ to 20 mag.

\section{Results }
Our sample of 1148 RRLab stars contains 288 thick disc stars with $1
< |z| < 4$ kpc, and also 860 halo stars with $|z| > 4$ kpc (see
Figure 3 of paper I). We use the equations introduced above section to
demonstrate the stellar number density and velocity distributions of the thick disk and halo.

\subsection{The thick disk profile}

Figure 1 demonstrates the stellar density map in the ${R_{\rm GC}}$--Z plane for the thick disk, the bin size is $2\times0.5$ kpc. We assume the thick disk has a shape of cylinder, and get the volume from the ${R_{\rm GC}}$ and Z (Z as a height). The thick disk has a range of $1< |z| < 4$ kpc in the vertical direction, and there is a gap between -1 and 1 kpc because of thin disk, RRLab stars are old and metal-poor stars. The tomographic map distributed in a Galactocentric distance range of 4.5--14.5 kpc. There are two ring areas in the Figure 1, the high density ring showed at the region of ${R_{\rm GC}}= 8$ -- $10.5$ kpc  and $Z= -1.5$ -- $-2.5$ kpc, while the relatively lower density ring showed up at the region of ${R_{\rm GC}}= 8$ -- $ 9.5$ kpc  and $Z= 2.2$ -- $ 2.8$ kpc (for the similar results see Newberg et al (2003), Juri$\acute{\rm c}$ et al. (2008) and Ivezi$\acute{\rm c}$ et al. (2008)). LAMOST survey covers different areas at different longitudes. Thus, these two density regions might be caused by the selection effect of LAMOST. From $Z= -1.5$ -- $- 4$ kpc the density distribution yield a heart shape between ${R_{\rm GC}}= 7$ -- $ 13$ kpc, and gradually decreases until ${R_{\rm GC}}= 14.5$ kpc (the outer edge). Combining our results of the thick disk and halo, we agree the conclusion of Liu et al. (2017) which is that the disk smoothly transit to the halo without any truncation (also see Liu et al. 2017).

The metallicity and velocity profiles of the thick disk RRLab stars are given in the upper and lower panels of Figure 2.
Most of the RRLab are metal-poor stars which have [Fe/H] around or lower than -1 dex while few of them with around 0. The red line in the upper panel of the figure is the mean value ([Fe/H]$=-1.02$) of all the thick disk tracers.  There are three stars which have the radial velocity $<$ -215 km ${\rm s}^{-1}$ around 7.5 kpc, while other the RRLab have the radial velocities higher than -210 km ${\rm s}^{-1}$ (see $V_{\rm FSR}$ distribution in the lower panel of Figure 2).

\subsection{The halo profile}

The number density of RRL has been on debate such as break or no break in the density distribution.
Ivezi$\acute{\rm c}$ et al. (2000) claimed the existence of a break in the density distribution in the halo
at ${R_{\rm GC}}\sim 50$ kpc by using 148 SDSS RRLs. However, Ivezi$\acute{\rm c}$ et al. (2004) and
Vivas \& Zinn (2006) have found no break until $\sim$ 60 or 70 kpc. A broken-power law has been considered
as a better number density profile for the RRLs by a number of works (e.g., Saha 1985; Sesar et al. 2007; Keller et al. 2008; Watkins et al. 2009; Akhter et al. 2012; Faccioli et al. 2014). Drake et al. (2013b) presented 1207 RRLs taken by the Caltalina Survey's Mount Lemmon telescopes, and found the number density out to 100 kpc with $\sim$70\% detection efficiency and a break appeared around 50 kpc (see the Figure 12 of Drake et al. (2013b)). They claimed their density profile is good agreement with the Watkins et al.(2009), and the different break is caused by the density enhanced by RRLs in the Sagittarius stream leading and trailing arms. We further analyze 860 RRLab
stars from Drake et al. (2013a) by combining LAMOST DR4 and SDSS DR8 data to show the density profile in the 9--50 kpc range. We consider a spherical averaged number density and fit by following formula,

\begin{equation}
n(R_{\rm GC}) = n_0 \left\{ \begin{array}{ll}
(\frac{R_0}{R_{\rm GC}})^{\alpha} & \textrm{if $R_{\rm min}<{R_{\rm GC}} < R_0$}\\
\\
(\frac{R_0}{R_{\rm GC}})^{\beta} & \textrm{if $R_0<{R_{\rm GC}} < R_{\rm max}$},
\end{array} \right.
\end{equation}
and derive the following values for the halo parameters : $n_0 = 0.35\pm0.18$ ${\rm kpc}^{-3}$, $R_0=21\pm2$ kpc, $\alpha=2.8\pm0.4$ and $\beta=4.8\pm0.4$ (see Figure 3). Watkins et al. (2009) gave their best results as: $(n_0,\,R_0,\,\alpha,\,\beta)=(0.26\,{\rm kpc}^{-3},\,23\,{\rm kpc},\,2.4\,,4.5)$. Faccioli et al. (2014) demonstrated 318 RRLs observed by Xuyi Schmidt telescope photometric survey, and obtained the density profiles by including and removing the possible Sagittarius RRLs. There is not a significant difference between their results with and without Sagittarius RRLs. Their spherical double-power model results are shown in their table 2, and the result with all RRLs are $n_0 = 0.42\pm0.16$ ${\rm kpc}^{-3}$, $R_0=21.5\pm2.2$ kpc, $\alpha=2.3\pm0.5$ and $\beta=4.8\pm0.5$. It is obvious that our result are strongly supports the results of Watkins et al. (2009) and Faccioli et al. (2014).

Figure 4 shows the velocity $V_{\rm FSR}$ distributions of 860 RRLab stars. In general, the radial velocities of RRLab stars are smoothly distributed with the distance, and a small part has a higher radial velocities at inner Galaxy, this may caused by the stream effect. We find the radial velocity dispersion at 50 kpc is $\sim$78 km ${\rm s}^{-1}$ (for more details also see paper I), which is smaller than $\sim$90 km ${\rm s}^{-1}$ derived by Cohen et al. (2017). We used 860 RRLab stars to measure the radial velocity dispersion for the halo based on the SDSS and LAMOST spectroscopic surveys, and Cohen et al. (2017) determined the radial velocity dispersion by only using 112 RRLs based on the moderate resolution spectra with Deimos on the Keck 2 Telescope.

\section{Conclusions}

In this work, we have investigated the density profiles and velocity distributions of the thick disk and halo of the Milky Way, based on 1148 RRLab variables with precise distances (7\% uncertainty) and reliable radial velocities (uncertainty $<$ 15 km $\rm s^{-1}$) presented in paper I. The 288 thick disk RRLab stars have the mean metallicity of [Fe/H]$=-1.02$. Despite three RRLab with the radial velocity lower than -215 km $\rm s^{-1}$, other disk tracers distributed in a region $>$ -210 km $\rm s^{-1}$. Our result shows that the edge of the thick disk is around 14.5 kpc. The halo of the Milky Way have been studying by using RRLab variables. Comparing to previous works, we present a larger sample (860) of halo RRLab variables with the mean metallicity of [Fe/H]$=-1.33$.
The stellar density distribution of the halo tracers can be well fitted by a broken power-law, and power law index of -2.8 for $<$ 21 kpc \& the index of -4.8 for $\geq$ 21 kpc. This density distribution is agreed by most of other works, especially by the works which used RRLs as the tracer. The radial velocity dispersion at 50 kpc is $\sim$78 km $\rm s^{-1}$, and few halo tracers show high radial velocities while that of others smoothly distributed.

\section*{Acknowledgements}

This work is supported by JSPS International Postdoctoral Fellowship of Japan (P17022, JSPS KAKENHI grant no. 17F17022), and also supported by National Natural Science Foundation of China under grant number 11390371 and 11233004.
Guoshoujing Telescope (the Large Sky Area Multi-Object Fiber Spectroscopic Telescope LAMOST) is a National Major Scientific Project built by the Chinese Academy of Sciences. Funding for the project has been provided by the National Development and Reform Commission. LAMOST is operated and managed by the National Astronomical Observatories, Chinese Academy of Sciences.




\begin{center}
  
 \textbf{ REFERENCES}
  \end{center}

Ablimit, I. \& Zhao, G. 2017, ApJ, 846, 10

Akhter, S., Da Costa, G. S. et al. 2012, ApJ, 756, 23

Bell, E. F., Zucker, D. B., Belokurov, V., et al. 2008, ApJ, 680, 295

Bensby, T. 2014, A\&A, 562, A71

Binney, J. J., \& Dehnen, W. 2010, \mnras, 403, 1829

Bland-Hawthorn, J. \& Gerhand, O. 2016, ARA\&A, 54, 529

Cacciari, C., \& Clementini, G. 2003, in Stellar Candles for the Extragalactic
Distance Scale, ed. D. Alloin \& W. Gieren (Lecture Notes in Physics, Vol.
635; Berlin: Springer), 105

Chaboyer, B. 1999, in Post-Hipparcos Cosmic Candles, ed. A. Heck \& F. Caputo
(Astrophysics and Space Science Library, Vol. 237; Dordrecht: Kluwer), 111

Cohen, G. J., Sesar, B., Bahnolzer, S., et al. 2017, arXiv:1710.01276v1

Cui, X-Q., Zhao, Y-H., Chu, Y-Q. et al., 2012, RAA, 12, 1197

Dambis, A. K., Berdnikov, L. N., Kniazev, A. Y. et al. 2013, \mnras, 435, 3206

Deason, A. K., Belokurov, V. \& Evans, N. W. 2011, MNRAS, 416, 2903

Deason, A. J., Belokurov, V.,  Evans, N. W. \& Johnston, K. V. 2013, \apj, 763, 113

Demarque, P., Zinn, R., Lee, Y-W. \& Yi, S. 2000, AJ, 119, 1398

Drake, A. J., Catelan, M., Djorgovski, S. G., et al. 2013a, \apj, 763, 32

Drake, A. J., Catelan, M., Djorgovski, S. G., et al. 2013b, \apj, 765, 154

Faccioli, L., Smith, M. C. et al. 2014, ApJ, 788, 105

Gillessen, S., Eisenhuaer, F., Trippe, S., et al. 2009, \apj, 692, 1075

Habing, H. 1988, A\&A 200, 40

Hawkins, K., Jofre, P, Masseron, T. \& Gilmore, G. 2015, MNRAS, 453, 758

Iorio, G., Belokurov, V., Erkal, D. et al. 2017, arXiv:1707.03833v2

Iverzi$\acute{\rm c}$, $\check{\rm Z}$. et al. 2000, AJ, 120, 963

Iverzi$\acute{\rm c}$, $\check{\rm Z}$. et al. 2004, in Clemens D., Shah R. Y., Brainerd T., eds,
APS Conf. Ser. Vol. 317, Milky Way Survey: The Structure and Evolution of Our Galaxy. Astron. Soc. Pac.,
 San Francisco, p. 179

Iverzi$\acute{\rm c}$, $\check{\rm Z}$., Sesar, B.,  Juri$\acute{\rm c}$, M. et al. 2008, ApJ, 684, 287

Juri$\acute{\rm c}$, M., Iverzi$\acute{\rm c}$, $\check{\rm Z}$., Brooks A. et al.
2008, ApJ, 673, 864

Keller, S. C., Murphy, S. et al. 2008, ApJ, 678, 851

Liu, C., Xu, Y., Wan, J-C. et al. 2017, RAA, 17, 96L

Miceli, A., Rest, A., Stubbs, C. W. et al. 2008, ApJ, 678, 865

Minniti, D., Saito, R., Alonso-Garcia, J., Lucas, P. \& Hempel, M. 2011, ApJ, 733, L43

Newberg, H. J., Yanny, B., Rockosi, C. et al. 2002, ApJ, 569, 245

Reid, M. J., Menten, K. M., Brunthaler, A. et al., 2014, \apj, 783, 130

Saha, A. 1985, ApJ 289, 310

Sandage, A. R. 1981, \apj, 248, 161

Sesar, B. et al., 2007, ApJ, 134, 2236

Sesar, B. et al., 2010, ApJ, 708, 717

Sesar, B. et al., 2011, ApJ, 731, 4

Smith, H. A. 1995, IrAJ 22, 228S

Watkins, L. L., Evans, N. W., Belokurov, V. et al. 2009, MNRAS, 398, 1757

Wetterer, C. J. \& McGraw, J. T. 1996, AJ, 112, 1046

Zhao, G., Chen, Y.-Q., Shi, J.-R. et al. 2006,
ChJAA, 6, 265

Zhao, G., Zhao, Y.-H., Chu, Y.-Q., Jing, Y.-P., Deng, L.-C. 2012,
RAA, 12, 723

\clearpage

\begin{figure}
\centering
\includegraphics[totalheight=3.5in,width=4.5in]{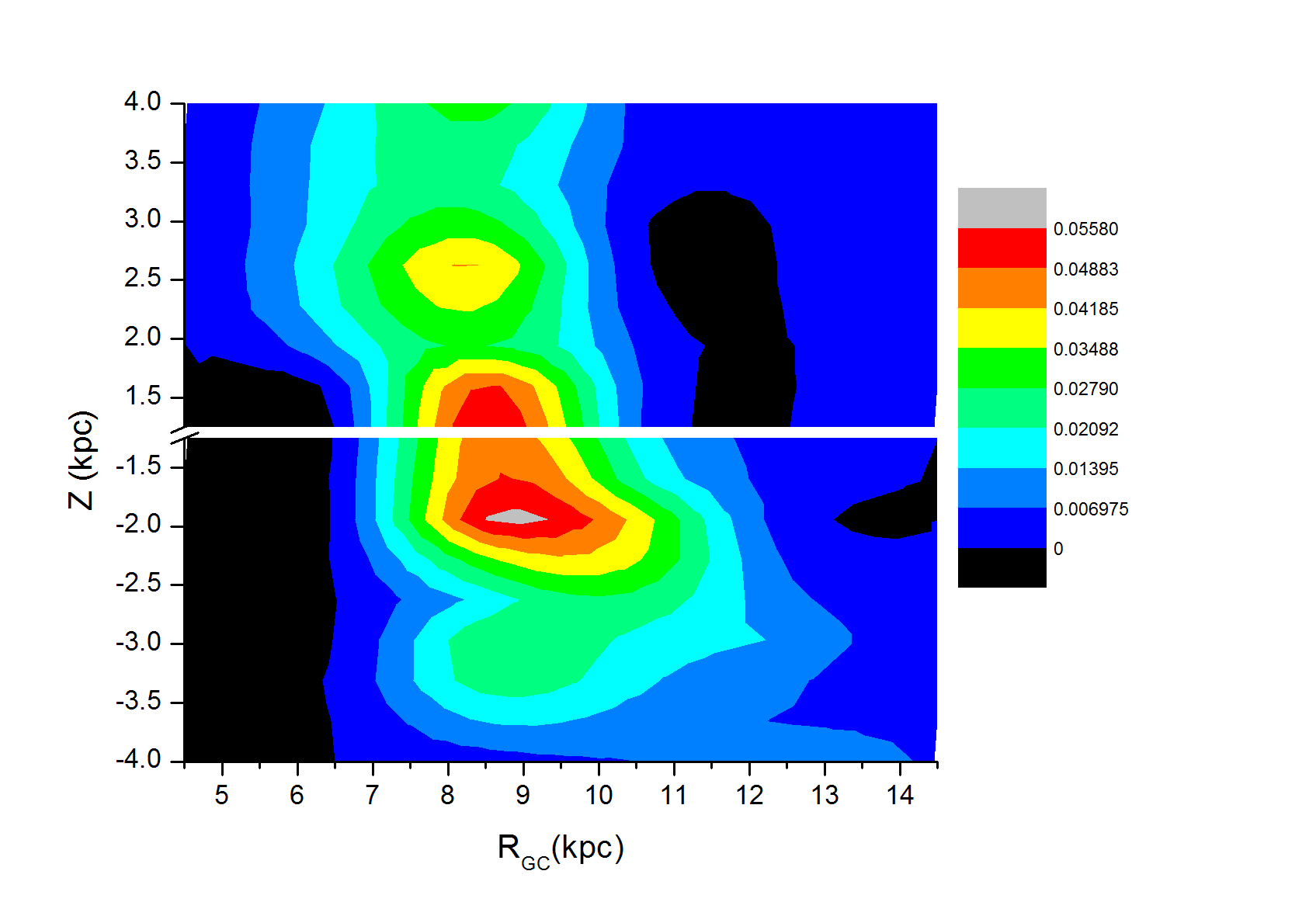}
\caption{The mean spatial density for RRLab stars. The color codes for each $R_{\rm GC}$--Z bin.
There is gap between -1 and 1 kpc, because the thick disk is considered as a disk of $1<|Z|<4$ kpc.}
\label{fig:1}
\end{figure}

\clearpage

\begin{figure}
\centering
\includegraphics[totalheight=3.2in,width=4.1in]{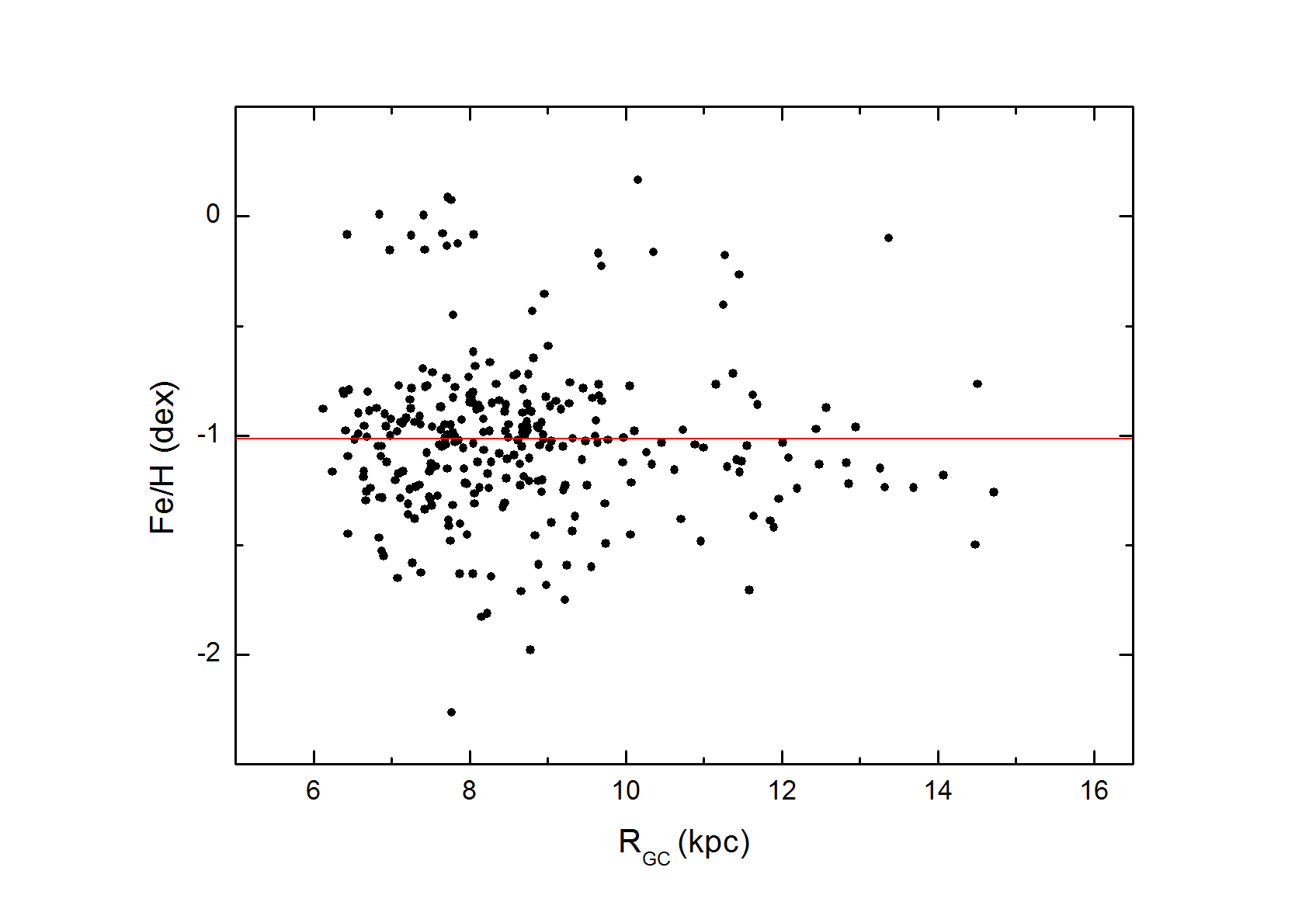}
\includegraphics[totalheight=3.5in,width=4.7in]{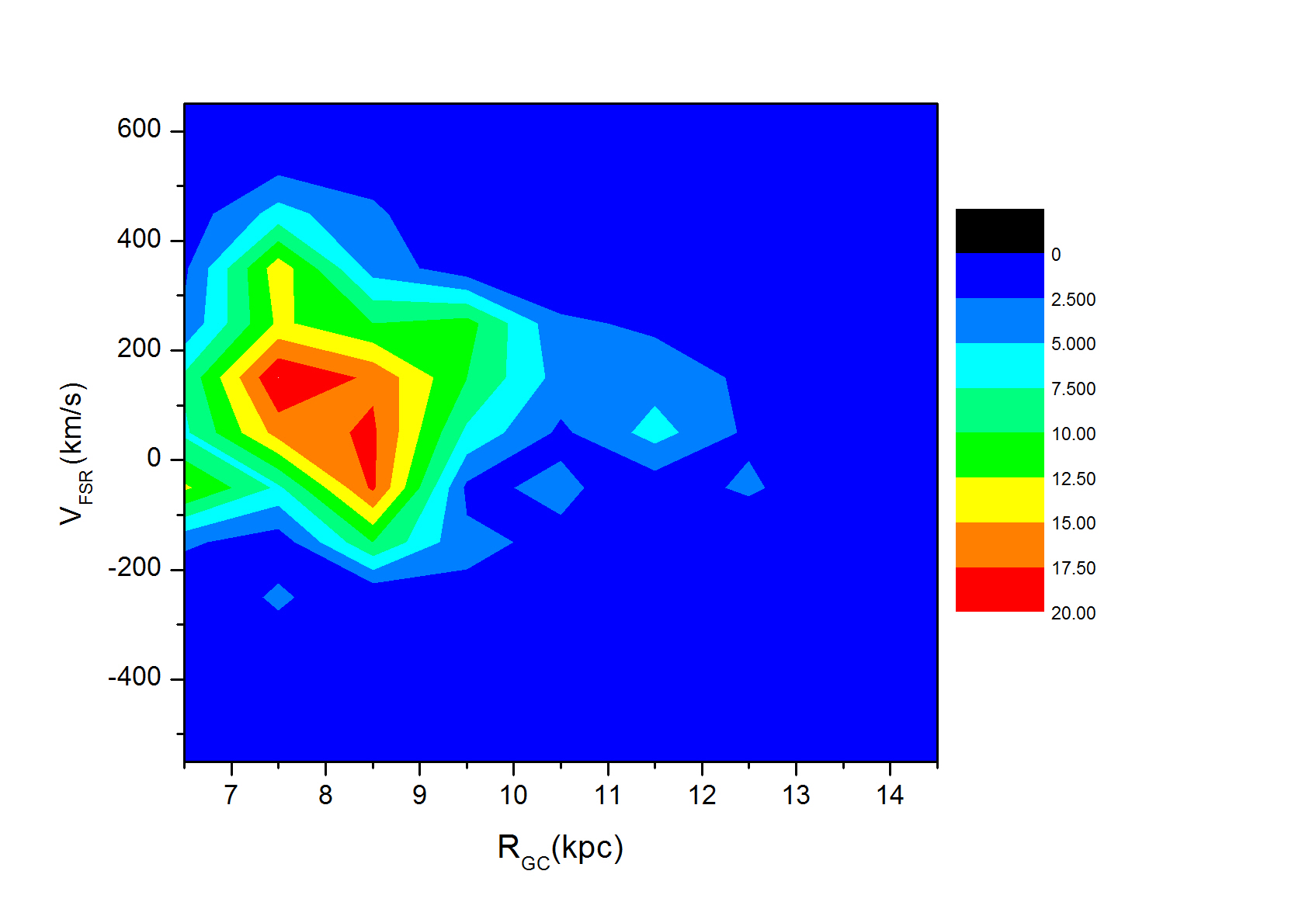}
\caption{The upper panel shows the metallicity distribution of the thick disk tracers (the red line represents the mean value), and the velocity distribution of the fundamental standard rests are shown in the lower panel.}
\label{fig:1}
\end{figure}

\clearpage

\begin{figure}
\centering
\includegraphics[totalheight=3.5in,width=4.5in]{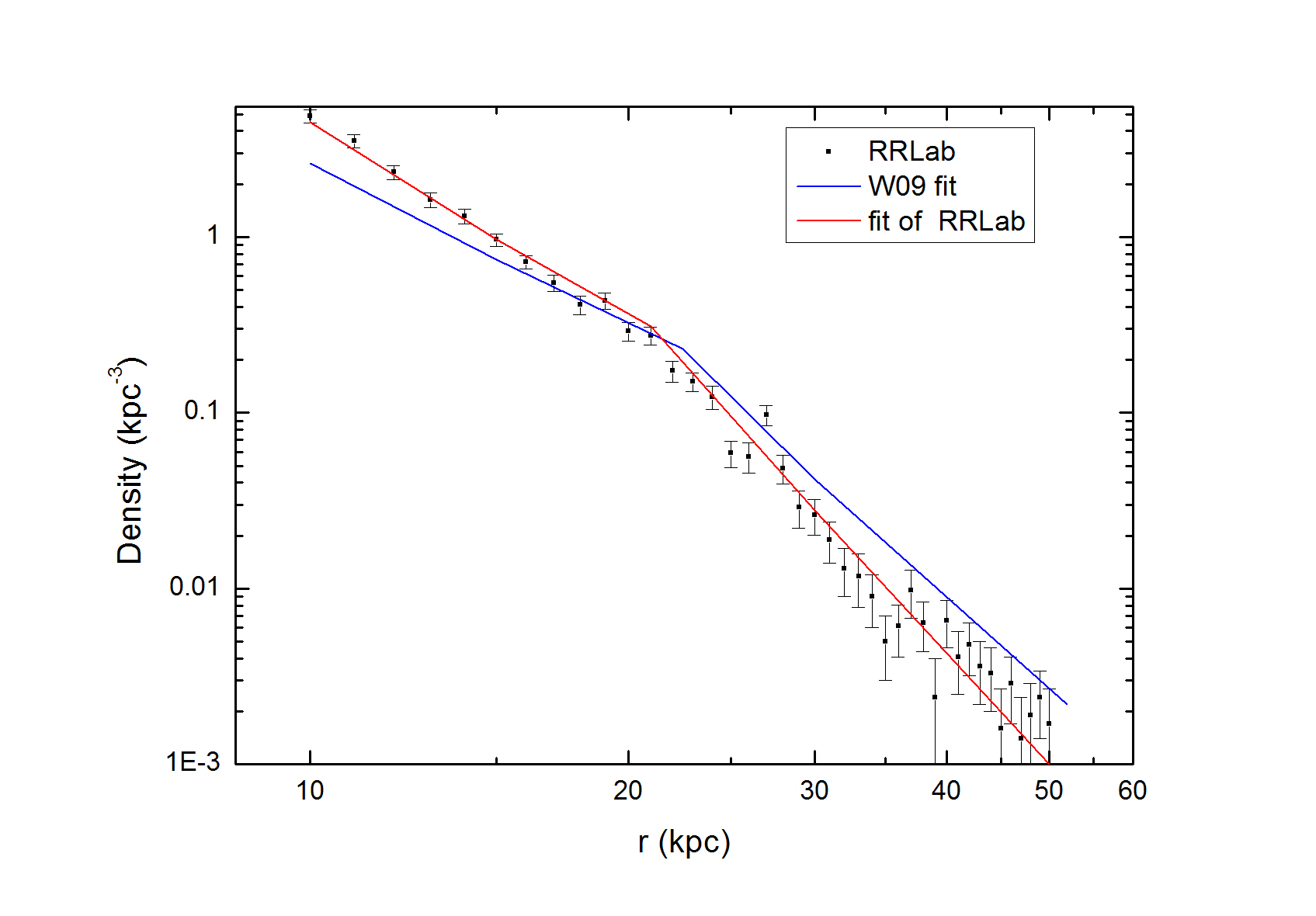}
\caption{The number density of the halo RRLab stars as a function of Galactocentric distance in a range of 9--50 kpc. The blue line is the fit result of Watkins et al. (2009), and the red line is the result derived in this work.}
\label{fig:1}
\end{figure}

\clearpage

\begin{figure}
\centering
\includegraphics[totalheight=3.2in,width=4.7in]{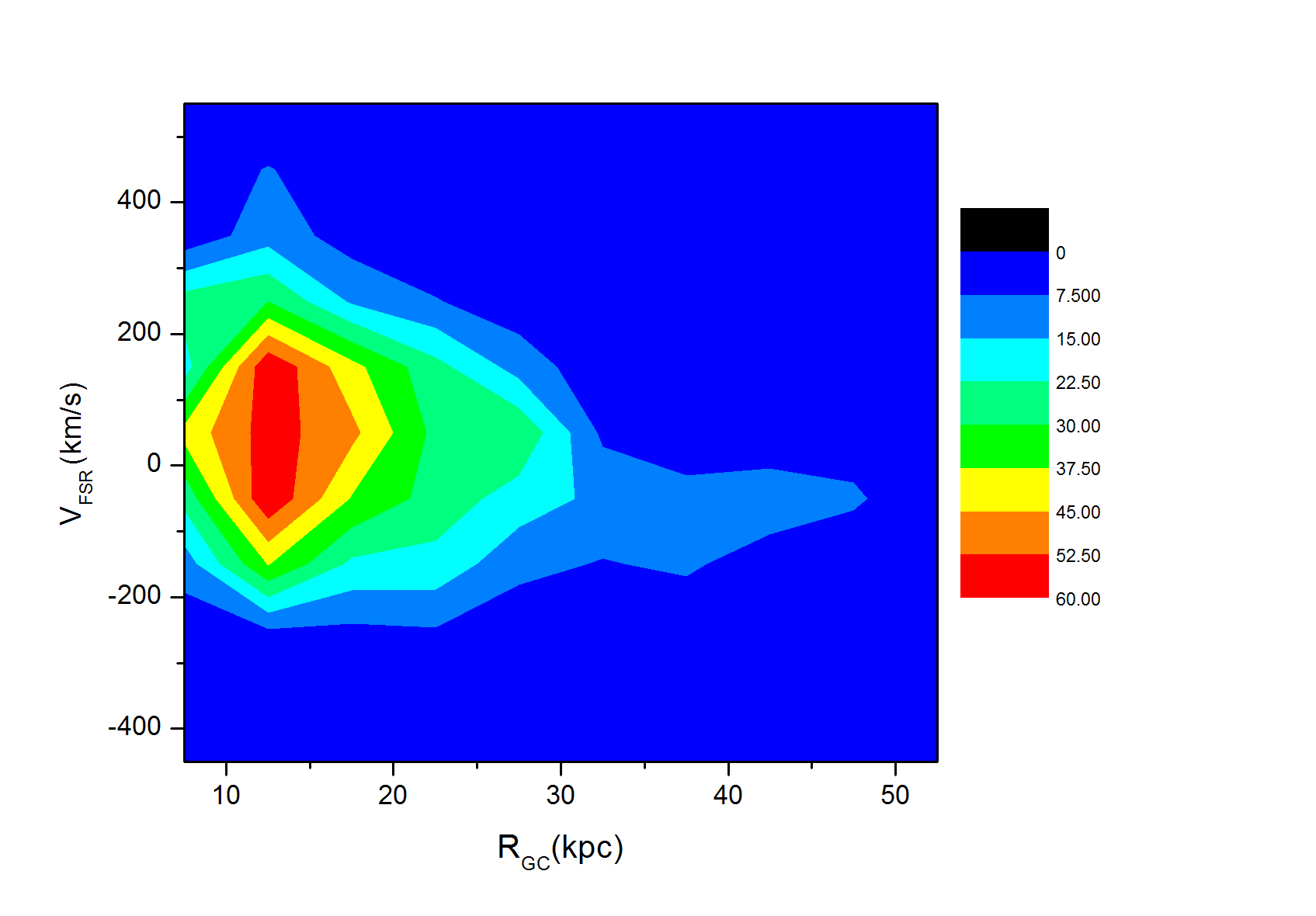}
\caption{The velocity distribution of the fundamental standard rests $V_{\rm FSR}$ of the halo RRLab stars.}
\label{fig:1}
\end{figure}

\clearpage



\end{document}